\documentclass[12pt]{article}

\usepackage{amsmath,amssymb,array,calc,rotating,epsfig,psfrag, amscd, datetime, comment, color, mathrsfs, ifthen}

\usepackage[
      colorlinks=true,
      linkcolor=darkblue,  
      urlcolor=blue,    
      filecolor=blue,     
      citecolor=red,
      ]{hyperref}
\usepackage[left=1cm,top=1cm,right=1cm,nohead]{geometry}
\usepackage{hyperref,slashed,empheq}
\usepackage{framed}
\usepackage{slashed}
\usepackage{simpler-wick}
\definecolor{cardinal}{rgb}{0.6,0,0}
\definecolor{darkgreen}{rgb}{0,0.5,0}
\definecolor{golden}{rgb}{0.92, 0.7, 0}
\definecolor{midnight}{rgb}{0, 0, 0.5}
\definecolor{darkblue}{rgb}{0.2, 0, 0.8}

\def\cN{{\cal N}}

\begin{document}
%%%%%%%%%%%%%%%%%%%%%%%%%%%%%%%%%%%%

%\begin{center}
%\fbox{\today,\ \currenttime}
%\end{center}

\vspace{0.5cm}
\begin{center}
\baselineskip=13pt {\LARGE \bf{On Gluing CFT Correlators to Higher-Spin Amplitudes
in AdS}}
 \vskip1.5cm 
Shailesh Lal
 \vskip0.5cm
\textit{Centro de Fisica do Porto e Departamento de Fisica e
  Astronomia \\ Faculdade de Ciencias da Universidade do
  Porto,\\ Rua do Campo Alegre 687, 4169-007 Porto, Portugal}\\
\vskip0.5cm
slal@fc.up.pt \\
\end{center}

\abstract{We demonstrate how three-point correlation functions of the free
scalar U(N) model involving two scalar operators and one spin-$s$ 
conserved current organize themselves
into corresponding AdS amplitudes involving two scalar and one spin-$s$ bulk to boundary
propagators, coupled via the bulk gauge invariant interaction vertex.
Our analysis relies on the general program advocated in hep-th/0308184
and some features of the embedding space formalism also play an important role.}

%%%%%%%%%%%%%%%%%%%%%%
\section{Introduction}
\label{sec:intro}
Higher-Spin/CFT dualities \cite{Sezgin:2002rt,Klebanov:2002ja,Gaberdiel:2010pz}, 
see \cite{Giombi:2016ejx,Gaberdiel:2012uj} for reviews,
have been of great interest since their discovery for a variety of reasons.
Firstly, since unitary representations of the Poincare group exist for arbitrarily high spins, it is
certainly of interest to enquire if consistent, interacting theories involving such fields may be constructed,
especially since a wide variety of no go theorems seem to preclude the existence of such theories
at first sight. While
we refer the reader to the reviews 
\cite{Bekaert:2010hw,Rahman:2015pzl} for a fuller account 
of the history of the field, 
it is important to especially note here 
that consistent, interacting theories involving higher-spin fields do exist in 
Anti-de Sitter space, and turn out to be of foundational importance to AdS/CFT duality. 
\footnote{We also note recent work on the construction of consistent interacting higher-spin theories in flat space
in \cite{Ponomarev:2016lrm,Ponomarev:2017nrr,Skvortsov:2018jea}, where the no-go theorems are
evaded by working in light-cone coordinates.}
Indeed, while the construction of the first such theories
\cite{Vasiliev:1990en}, reviewed in \cite{Didenko:2014dwa},
significantly predates 
the proposal of AdS/CFT duality,
the duality also almost seems to require the existence of 
higher-spin theories in AdS \cite{Sundborg:1999ue,HaggiMani:2000ru,Sundborg:2000wp}. We briefly
recollect the elements of this argument.

For definiteness, let us consider the original duality between Type IIB string theory on AdS$_5\,\times$ S$^5$
and $\cN=4$ SYM defined on the boundary of the AdS$_5$ 
\footnote{Here and throughout, we shall be concerned with the Euclidean
version of the duality, with the AdS$_{d+1}$ taken to be the 
Poincare patch, with boundary $\mathbb{R}^d$. A significant
role is played by the embedding space formalism, 
in which these spaces are taken to lie in $\mathbb{M}^{1,d+1}$. Details
of this formalism are briefly presented in Appendix \ref{app:embedding} and reviewed
more extensively in \cite{Costa:2014kfa,Bekaert:2010hk,Sleight:2016hyl,Sleight:2017krf,Rychkov:2016iqz}.}
\cite{Maldacena:1997re}. The boundary parameters $\lambda$ and $N$ are 
related to the bulk parameters string tension $\alpha'$, 
Planck length $\ell_P$, and string length $\ell_s$ via
\begin{equation}
N^2 \sim {\left(\ell_{\text{AdS}}\over\ell_P\right)^{8}}, 
\lambda \sim {{\ell_{AdS}}^4\over \left(\alpha^\prime\right)^2}\,.
\end{equation}
From the above, we see that the large-$N$ limit on the boundary maps to a weakly coupled (semiclassical) 
limit on the bulk, while the $\lambda=0$ limit on the boundary sets the string tension to zero in the bulk.
Hence, a free planar limit of the CFT maps to a classical tensionless string in AdS. It has long been expected
that many hidden symmetries of string theory might be manifest in this limit,
much as the massless limit of QFTs makes their hidden symmetries manifest. 
Carrying over intuition from tensile string theory in flat space, the tensionless limit 
is also precisely where we expect massless higher-spins to appear in the string spectrum. 
We shall shortly review how AdS/CFT furnishes much sharper reasons to expect the same. 
In fact, a careful analysis, see e.g. \cite{Beisert:2003te,Beisert:2004di} 
indicates that higher-spin symmetry would indeed play an important
role in this limit of the duality.
Finally, though the above expressions were for one specific instance of AdS/CFT, 
the arguments presented here are expectedly
generic to all known cases of the duality. We particularly note the recent developments
\cite{Eberhardt:2018ouy} where these 
expectations have been very explicitly realized for the tensionless
string in AdS$_3$.

Moreover, in this limit both the bulk and the boundary theories admit a weak coupling expansion, so 
perturbative tests of the duality are possible, and indeed have been carried out for the case of `pure'
higher-spin/CFT dualities \cite{Giombi:2009wh,Giombi:2010vg,Chang:2011mz}, see also
\cite{Colombo:2012jx,Didenko:2012tv,Didenko:2013bj}. 
Further, holographic reconstructions of the interaction vertices of the
AdS higher-spin theory have been explored 
\cite{Sleight:2016dba,Bekaert:2014cea,Bekaert:2015tva,Bekaert:2016ezc}, 
as reviewed for instance in \cite{Sleight:2016hyl,Sleight:2017krf}.

Given this situation, it is natural to enquire if one may use this window to gain some insight into
the underlying mechanics of AdS/CFT duality. In particular, since both the bulk and boundary expressions
are perturbative, and presumably admit at least some degree of simultaneous control, 
one may wish to systematically rewrite expressions on one side of the duality as their counterparts 
on the other side. Specifically, we ask if we may take a CFT correlation
function, computable by Wick contractions, and systematically rewrite it as the sum of Witten diagrams 
in AdS? It would clearly be very interesting to identify a precise mechanism by which this would happen.
For one, the original proposal of \cite{Maldacena:1997re} 
seems to rely deeply on dynamics of D-branes in string 
theory, supersymmetry, and open-closed string duality. 
However, the actual computational prescription for comparing bulk
and boundary theories in the classical bulk limit makes little to no reference to a string theory, using 
instead the symmetries of the bulk and boundary \cite{Gubser:1998bc,Witten:1998qj} as a guiding principle. 
Identifying the precise mechanism by which bulk and boundary theories organize into each other 
could shed new light on the interplay
between these two seemingly disparate features that underpin the duality.

A general prescription for carrying this out, inspired by open-closed string duality, was proposed in 
\cite{Gopakumar:2003ns,Gopakumar:2004qb,Gopakumar:2004ys,Gopakumar:2005fx}.
Schematically, one starts with the CFT correlator written in a slight variation of 
Schwinger parametrization and carries out a change of variables on these parameters. 
For the all-scalar three-point function \footnote{The three-point function is a natural starting
point for this program, as there is only one dual Witten diagram to organize the CFT correlator into.}
, the resulting expression quite naturally
became a product of three scalar bulk to boundary propagators meeting at a bulk point, which was integrated
over \cite{Gopakumar:2003ns}. The boundary coordinates of the bulk point arose from momentum conservation
in the correlator, while the extra holographic coordinate was obtained from the Schwinger parameters. 
These facts may be quickly reviewed by following our equations \eqref{cft3pt00s} 
through to \eqref{scalar gluing} all the while setting $s=0$ there. In what follows, we shall follow 
the nomenclature of \cite{Gopakumar:2003ns} and refer to their proposed prescription as `gluing'.

In this paper, we shall extend this analysis to the three point function of two scalar operators 
with one conserved current of arbitrary spin. 
\footnote{Gluing for some `low spin' three-point correlators
was previously studied in \cite{Mamedov:2005uw,Konyushikhin:2007zz}.}
Even at the outset one may notice several potential subtleties.
On the AdS side, the explicit form of the interaction vertex may be changed on integration by parts, and the form 
of the spinning propagator is also gauge dependent. Additionally, there is the freedom of doing
field redefinitions in the bulk. Having glued the CFT correlation 
function as per \cite{Gopakumar:2003ns}, how may one recognize it as having the form of the appropriate
AdS amplitude? In this regard, the embedding space formalism \cite{Fronsdal:1978vb} 
turns out to be especially useful, which
should perhaps not surprise us, given its great utility in studying 
higher-spin fields and their interactions, see e.g. 
\cite{Bekaert:2010hk,Joung:2011ww,Joung:2012fv,Manvelyan:2012ww,Joung:2013doa,
Metsaev:1995re,Metsaev:1997nj,Fotopoulos:2006ci,Alkalaev:2009vm},
as well as correlators in conformal field theory, see e.g. 
\cite{Costa:2011mg,Costa:2011dw,Costa:2016xah} and \cite{Rychkov:2016iqz} for 
an introduction to the formalism for CFTs.

We shall now turn to a brief account of the free CFT and its spectrum and holographic dual, then in Section \ref{sec:cft}
obtain the form of CFT correlator in embedding space after gluing. We then show in Section \ref{sec:ads} that this 
indeed has the structure we expect from the AdS amplitude. We then conclude.

A final note on conventions. We work Euclidean signature, using coordinates $\vec{x}$ on 
the $d$ dimensional boundary, and coordinates $\mathbf{x}\equiv\left(t,\vec{w}\right)$ on 
the Poincare patch of AdS$_{d+1}$. The usual holographic coordinate $z$ on the Poincare
patch is related to $t$ via $z=\sqrt{t}$. The AdS$_{d+1}$ metric in our coordinates is
\begin{equation}
ds^2 = {dt^2\over t^2}+{1\over t}\,d\vec{w}^2\,,\qquad 
\sqrt{g}\,d^{d+1}\mathbf{x}={dt\over t^{d/2+1}}\,d^dw\,.
\end{equation}
We shall often drop the overhead arrows.
Further, since we work extensively with symmetric tensors, we
shall repackage indices in terms of a polarization vector. For example,
a spin-$s$ operator in the boundary will be written as 
$J_s\left(x,z\right)=J_{\mu_1\ldots\mu_s}\left(x\right)z^{\mu_1}\ldots z^{\mu_s}$.
The polariation vector in the bulk will be denoted by $u$. Small letters denote intrinsic
quantities, and capital letters denote quantities in embedding space.
\subsection{The Free CFT on the Boundary and the Interacting Theory in the Bulk} 
We shall most concretely be working with the free $U(N)$ vector model with fundamental scalars
in $\mathbb{R}^{d}$, i.e.
\begin{equation}\label{cft un vector}
S=\int\,d^dx\,\varphi^{i\,*}\left(x\right)\,\partial^2\,\varphi_i\left(x\right)\,,
\end{equation}
where we shall consider the set of single trace conformal primary operators involving two insertions
of the field $\varphi_i$. This is given by \cite{Craigie:1983fb}
\footnote{Higher-spin conserved currents in Minkowski space were also
constructed in \cite{Berends:1985xx,Anselmi:1999bb,Vasiliev:1999ba,Konstein:2000bi}.
}
\begin{equation}\label{js}
\begin{split}
J_s\left(x,z\right)& = \varphi^*\left(x\right)\,f^{(s)}
\left(z\cdot\overleftarrow{\partial},
z\cdot\overrightarrow{\partial}
\right)
\,\varphi\left(x\right)\,,\\
&\quad f^{(s)}\left(u,v\right)
=\left[2^{s}\tfrac{\left(\Delta-1\right)_s\left(\Delta-1\right)_{2s}}
{\Gamma\left(s+1\right)}\right]^{-1/2}
\left(u+v\right)^s C^{\left(\Delta/2 -1\right)}_s\left(u-v\over u+v\right)\,,
\end{split}
\end{equation}
where $z^2=0$ encodes tracelessness of the primary, $s$ is an
arbitrary positive integer, corresponding to the spin of the primary, and
$C^{\left(\Delta/2 -1\right)}_s$ is a
Gegenbauer polynomial. 
Here, and throughout, $\Delta=d-2$, 
which is the dimension of the scalar primary $\varphi^{*\,i}\,\varphi_i$.
The normalization constant is from \cite{Sleight:2016dba}, with a relative $\sqrt{2}$
since we have a complex scalar instead of a real one,
and is chosen such that 
\begin{equation}
\left\langle\,J_s\left(x_1,z_1\right)\,J_s\left(x_2,z_2\right)\, \right\rangle
={1\over{x_{12}\,^{\Delta+2s}}}\left(z_1\cdot z_2-
2\,{z_1\cdot x_{12}\,z_2\cdot x_{12}\over x_{12}^2}\right)^s\,.
\end{equation}
For most computations done in this paper it is usually more convenient to 
utilize the series expansion of $f^{(s)}$
and instead write
\begin{equation}\label{fs}
f^{(s)}\left(u,v\right)
=n_s^{-1/2}\,\sum_{k=0}^s {\left(-1\right)^k\over k!\left(s-k\right)!
\left(k+\tfrac{d-4}{2}\right)!\left(s-k+\tfrac{d-4}{2}\right)!}\,u^k\,v^{s-k}\,,
\end{equation}
where 
\begin{equation}\label{ns}
n_s = \frac{(-1)^s 2^{\Delta +3 s-2} \Gamma \left(s+\frac{\Delta -1}{2}\right)}
{\sqrt{\pi }\,s!\, \Gamma \left(\frac{\Delta }{2}\right)^2 
\Gamma \left(s+\frac{\Delta }{2}\right) 
\Gamma (s+\Delta -1)}\,.
\end{equation}
In particular, note that $n_0=\Gamma\left(\tfrac{\Delta}{2}\right)^{-4}$, so that
$f^{(0)}\left(u,v\right)=1$.
One may check by explicit series expansion 
that \eqref{js} and \eqref{fs} define the same currents, as we have done in
Appendix \ref{app:current normalization}.
We also note that
with the exception of the $s=0$ operator, these primaries all saturate the unitarity
bound \cite{Mack:1975je,Minwalla:1997ka} and hence are conserved currents. 

As should be apparent, for the vector model this completely exhausts
the set of single trace conformal primaries. For a matrix valued scalar, this would be a subset,
the full sprectrum comprising of long multiplets of the conformal algebra as well as 
representations of mixed symmetry. Explicit enumerations of the spectrum are (partially) available
in \cite{Bae:2016rgm,Bae:2016hfy}.
However, even in the matrix case, 
our analysis would continue to apply for that subset and should also extend
to scalar bilinear operators carrying additional internal symmetry indices, 
as in $\cN=4$ SYM. A few examples of
such primaries in $\cN=4$ SYM are available in Table 1 of \cite{Sundborg:2000wp}, see 
\cite{Beisert:2003te,Beisert:2004di} for more details regarding this spectrum.

As a final observation
regarding the CFT, we note that though the theory is free, the correlation functions of the operators
$J_s$ contain non-zero connected components, as is well known and is widely available
in the literature. The connected component of
$\left\langle\,J_0\,J_0\,J_s\,\right\rangle$,
first evaluated in \cite{Diaz:2006nm},
has also been explicitly evaluated below and 
is proportional to ${1\over\sqrt{N}}$. Standard AdS/CFT arguments, 
based on the dictionary \cite{Gubser:1998bc,Witten:1998qj} therefore 
lead us to expect a dual theory \cite{Vasiliev:1990en,Vasiliev:2003ev} 
with an AdS$_{d+1}$ vacuum with an infinite tower of 
higher spins weakly coupled to each other. As mentioned before, it is remarkable
that though the CFT \eqref{cft un vector} is not embedded in string theory,
a well defined AdS dual has been explicitly proposed 
in many cases \cite{Sezgin:2002rt,Klebanov:2002ja}, note also the low-dimensional
instance of \cite{Gaberdiel:2010pz}, and the dualities
seem to obey all the standard features of a stringy AdS/CFT duality.

%%%%%%%%%%%%%%%%%%%%%%
\section{The CFT Three-Point Functions}\label{sec:cft}
The three-point function 
$\left\langle J_0\left(x_1\right) J_0\left(x_2\right) J_s\left(x_3,z\right)\right\rangle$
may readily be computed in the CFT by Wick contractions and determined to be 
\footnote{
We refer the reader to \cite{Osborn:1993cr,Costa:2011mg,Stanev:2012nq,Zhiboedov:2012bm} 
where two and three-point 
functions of arbitrary-spin conserved currents
have been extensively studied. We also note
previous work 
\cite{Giombi:2011rz,Colombo:2012jx,Didenko:2012tv,Gelfond:2013xt} 
in $d=3$.
Explicit expressions for these current correlators in the 
$O(N)$ vector model were obtained in
\cite{Diaz:2006nm,Sleight:2016dba}.}
\begin{equation}\label{cft3ptstandard}
\left\langle J_0\left(x_1\right) J_0\left(x_2\right) J_s\left(x_3,z\right)\right\rangle
=g\,{1\over \left(x_{12}\right)^\Delta \left(x_{23}\right)^\Delta \left(x_{31}\right)^\Delta}
\left[ {z\cdot x_{13}\over x_{13}^2}-{z\cdot x_{23}\over x_{23}^2}\right]^s\,.
\end{equation}
This is the form of the expression as expected from conformal symmetry.
%Bose symmetry requires the right hand side to be symmetric under 
%the exchange of $x_1$ and $x_2$ while the term
%in square brackets is anti-symmetric under this exchange. 
%This requires that the correlator vanishes
%for all odd spins $s$.
For the purposes of gluing into the AdS amplitude, we note that there
is an alternate form of this correlator.
%An equivalent way of writing this correlator, and the one we shall work
%most closely with, is
\begin{equation}\label{cft3pt00s}
\begin{split}
\left\langle J_0\left(x_1\right)J_0\left(x_2\right)
J_s\left(x_3,z\right)\right\rangle&=
\left(1+\left(-1\right)^s\right)
\int d^dw
\int_0^\infty {d\tau\over\tau^{d/2+1}}\,\tau^3\,
\int_0^1\mathrm{d}^3\alpha\,\times
\\&\times
f^{(s)}\left(\tfrac{z\cdot x_{13}}{2\tau\alpha_2},
\tfrac{z\cdot x_{23}}{2\tau\alpha_1}\right)
\tfrac{e^{-\sum{\left(x_i-w\right)^2\over 4\tau\alpha_j\alpha_k}}}
{\tau^{3d/2}\left(\alpha_1\alpha_2\alpha_3\right)^d}\,.
\end{split}
\end{equation}
Here, and in the rest of the paper, the measure
\begin{equation}
\mathrm{d}^3\alpha\equiv d\alpha_1\,d\alpha_2\,d\alpha_3\,
\delta\left(\Sigma_i\alpha_i-1\right)\,.
\end{equation}
While \eqref{cft3ptstandard} follows straightforwardly by computing the 
given correlator in the theory \eqref{cft un vector} by
Wick contractions
in position space,
it is less obvious that \eqref{cft3pt00s} is also an expression for the correlator.
%\footnote{The reader may easily check, as detailed in Appendix \ref{app:standard}, that 
%doing the $w$ integral explicitly reduces \eqref{cft3pt00s} to \eqref{cft3ptstandard}. 
%However, since we intend to show that the CFT correlator may be rewritten as an AdS
%amplitude, where $\vec{w}$ will play the role of boundary coordinates of the bulk AdS point
%to be integrated over, this argument would in some sense assume part of the conclusion.}
In practice \eqref{cft3pt00s} may most readily be arrived at by starting with the corresponding
expression in momentum space, and Fourier transforming to position
space. Carrying out this analysis to get a form suitable for gluing requires
some care, and we provide details below.
\subsection{The Three-Point Function in Momentum space}
We now
Equation \eqref{cft3pt00s}, which is the essential starting-point for gluing 
into the corresponding AdS amplitude. 
We closely follow the approach of \cite{Gopakumar:2003ns},
starting with the momentum space representation of the amplitude 
and then Fourier transforming to position space.
However, we work directly with the second quantized version of the theory rather than attempting to
develop a worldline description incorporating vertex operators for higher-spin currents. It would be
interesting to develop such a formalism, expectedly using \cite{Bekaert:2010ky,Bonezzi:2017mwr} 
as a starting point.
The spin-s conserved current in position space is
\begin{equation}
J\left(x,z\right)= \varphi^*\left(x\right)
f^{(s)}\left(z\cdot\overleftarrow{\partial},
z\cdot\overrightarrow{\partial}
\right)\varphi\left(x\right)\,,
\end{equation}
where $z^2=0$ and $f^{(s)}$ was defined in \eqref{fs}. 
We now take the Fourier transform of this equation to momentum space.
We have the following conventions. 
\begin{equation}
\begin{split}
f\left(x\right) = \int {d^dk\over\left(2\pi\right)^{d/2}}
\,e^{i\,k\cdot x} &f\left(k\right)\,,
\quad
f\left(k\right) = \int {d^dx\over\left(2\pi\right)^{d/2}}
\,e^{-i\,k\cdot x} f\left(x\right)\,,
\\&\delta^{(d)}\left(p\right) 
= \int {d^dw\over\left(2\pi\right)^d}\,e^{-iw\cdot p}\,.
\end{split}
\end{equation}
Then
\begin{equation}
J_s\left(p,z\right) 
=i^{s+2d}\,\int {d^dk\over\left(2\pi\right)^{d/2}}\,
\varphi^*\left(k\right)
f^{(s)}\left(z\cdot k,
z\cdot\left(p-k\right)\right)
\varphi\left(p-k\right)
\,,
\end{equation}
and the 3-point function in momentum space is then given by
\begin{equation}
\begin{split}
&\left\langle J_0\left(p_1\right)J_0\left(p_2\right)
J_s\left(p_3,z\right)\right\rangle= i^{s+2d}
\int \prod_{i=1}^3\,{d^dk_i\over\left(2\pi\right)^{d/2}}\,
f^{(s)}\left(z\cdot k_3,
z\cdot\left(p_3-k_3\right)\right)\times
\\&\qquad\qquad\times
\left\langle
\varphi^*\left(k_1\right)\varphi\left(p_1-k_1\right)
\varphi^*\left(k_2\right)\varphi\left(p_2-k_2\right)
\varphi^*\left(k_3\right)\varphi\left(p_3-k_3\right)
\right\rangle\,.
\end{split}
\end{equation}
Two sets of Wick contractions contribute to the connected
part of this correlator. 
\footnote{
We use the basic Wick contraction
\begin{equation}
\wick{\c1\varphi^{*}\left(k_1\right) \c1 \varphi\left(k_2\right)}
={4\pi^{d/2}\over\Gamma\left(\Delta\over 2\right)}
\delta^{d}\left(k_1+k_2\right)\,{1\over k_1^2}\,,
\end{equation}
obtained by
Fourier transforming the position space expression
\begin{equation}
\wick{\c1\varphi^{*}\left(x_1\right) \c1 \varphi\left(x_2\right)}
={1\over\Gamma\left(\Delta\over 2\right)}
\int_0^\infty {dt\over t}\,t^{\Delta} e^{-t x_{12}^2}\,.
\end{equation}
}
These are will be denoted by $\left\langle\ldots\right\rangle_{(1,2)}$
respectively. We find that
\begin{equation}\label{doublebracket00s}
\left\langle J_0\left(p_1\right)J_0\left(p_2\right)
J_s\left(p_3,z\right)\right\rangle_{(i)}=
{\left(-1\right)^d\over\left(2\pi\right)^{3d/2}}
\left[{4\pi^{d/2}\over\Gamma\left(\Delta\over 2\right)}\right]^3
\left\langle\left\langle J_0\left(p_1\right)J_0\left(p_2\right)
J_s\left(p_3,z\right)\right\rangle\right\rangle_{(i)}\,,
\end{equation}
where
\begin{equation}
\left\langle\left\langle J_0\left(p_1\right)J_0\left(p_2\right)
J_s\left(p_3,z\right)\right\rangle\right\rangle_{(1)}=
i^s\delta\left(\Sigma_k\,p_i\right)
\int d^dk\,
{f^{(s)}\left(z\cdot k,
z\cdot \left(p_3-k\right)\right)
\over\left(k+p_1\right)^2\left(k-p_3\right)^2\,k^2}\,,
\end{equation}
and
\begin{equation}
\left\langle\left\langle J_0\left(p_1\right)J_0\left(p_2\right)
J_s\left(p_3,z\right)\right\rangle\right\rangle_{(2)}=
i^s\delta\left(\Sigma_k\,p_i\right)
\int d^dk\,
{f^{(s)}\left(z\cdot k,z\cdot \left(p_3-k\right)\right)
\over \left(k-p_3\right)^2\left(k+p_2\right)^2\,k^2}\,.
\end{equation}
We simplify the denominator using Schwinger parametrization.
In the first Wick contraction,
\begin{equation}
{1\over\left(k+p_1\right)^2\left(k-p_3\right)^2\,k^2}
=\int_0^\infty\,d^3\tau\,
e^{-\tau_1\left(k-p_3\right)^2 -\tau_2\,k^2 -\tau_3\left(k+p_1\right)^2} \,.
\end{equation}
We now change integration variables to $\left(\tau,\alpha_i\right)$ 
where $\tau_i=\tau\alpha_i$
and $d^3\tau = \tau^2\,d\tau\,\mathrm{d}^3\alpha$
where $
\mathrm{d}^3\alpha = d\alpha_1\,
d\alpha_2\,d\alpha_3\, \delta\left(\Sigma_i\,\alpha_i-1\right)\,.$
We therefore find
\begin{equation}
\begin{split}
{1\over\left(k+p_1\right)^2\left(k-p_3\right)^2\,k^2}
= \int {d\tau\over\tau}\,\tau^3\int \mathrm{d}^3\alpha
e^{-\tau
\left(\ell^2
+\alpha_3\alpha_2\,p_1^2
+\alpha_1\alpha_2\,p_3^2
+\alpha_1\alpha_3\,p_2^2\right)}\,,
\end{split}
\end{equation}
where $\ell = k + \alpha_3\,p_1-\alpha_1 p_3$, and
we used $\alpha_1+\alpha_2+\alpha_3 = 1$.
As a result, 
\begin{equation}
\begin{split}
&\left\langle\left\langle J_0\left(p_1\right)J_0\left(p_2\right)
J_s\left(p_3,z\right)\right\rangle\right\rangle_{(1)}=i^s
\delta\left(\Sigma_k\,p_i\right)
\int d^d\ell\,
\int {d\tau\over\tau}\,\tau^3\int \mathrm{d}^3\alpha
e^{-\tau
\left(\ell^2
+\Sigma\alpha_i\alpha_j\,p_k^2\right)}
\times
\\&\qquad\times
f^{(s)}\left(z\cdot \left(\ell-\alpha_3p_1+\alpha_1p_3\right),
z\cdot \left(-\ell+\alpha_3p_1+\left(1-\alpha_1\right)p_3\right)\right)
\,.
\end{split}
\end{equation}
$f^{(s)}$ is polynomial in its arguments. Terms odd in $\ell$ integrate
to zero because they are odd functions. Terms even in $\ell$ integrate 
to zero because the $\ell$ integral is proportional to appropriately symmetrized
combinations of $\eta^{\mu\nu}$, and $z^2=0$. 
Hence only the $\ell^0$ term contributes non-vanishingly
to the integral, and may be evaluated using
the standard Gaussian integral formula
$\int d^d\ell\, e^{-\tau\ell^2} = \left(\pi\over\tau\right)^{d/2}$.
We also have the identities
\begin{equation}
\begin{split}
\alpha_3p_1+&\left(1-\alpha_1\right)p_3 = 
-\alpha_3\left(p_2+p_3\right)+\left(1-\alpha_1\right)p_3
\\&=-\alpha_3 p_2+\left(1-\alpha_1-\alpha_3\right)p_3
=-\alpha_3 p_2+\alpha_2 p_3\,.
\end{split}
\end{equation}
Hence
\begin{equation}
\begin{split}
\left\langle\left\langle J_0\left(p_1\right)J_0\left(p_2\right)
J_s\left(p_3,z\right)\right\rangle\right\rangle_{(1)}&=i^s\pi^{d/2}
\delta\left(\Sigma_k\,p_i\right)\,
\int {d\tau\over\tau^{d/2+1}}\,\tau^3\int \mathrm{d}^3\alpha\,
e^{-\tau\sum\alpha_i\alpha_j\,p_k^2}
\times
\\&\times
f^{(s)}\left(z\cdot \left(\alpha_1p_3-\alpha_3p_1\right),
z\cdot \left(\alpha_2p_3-\alpha_3p_2\right)\right)
\,.
\end{split}
\end{equation}
The sum in the exponent above runs over values $i\neq j\neq k$.
The second set of Wick contractions may be evaluated exactly analogously
and we obtain
\begin{equation}
\begin{split}
\left\langle\left\langle J_0\left(p_1\right)J_0\left(p_2\right)
J_s\left(p_3,z\right)\right\rangle\right\rangle_{(2)}&=i^s\pi^{d/2}
\delta\left(\Sigma_k\,p_i\right)
\int {d\tau\over\tau^{d/2+1}}\,\tau^3\int \mathrm{d}^3\alpha\,
e^{-\tau\sum\alpha_i\alpha_j\,p_k^2}
\times
\\&\times
f^{(s)}\left(z\cdot \left(\alpha_2p_3-\alpha_3p_2\right)
,z\cdot \left(\alpha_1 p_3-\alpha_3 p_1\right)\right)
\,.
\end{split}
\end{equation}
Since $f^{(s)}\left(y,x\right)=
\left(-1\right)^sf^{(s)}\left(x,y\right)$,
we have
\begin{equation}
\begin{split}
&\left\langle\left\langle J_0\left(p_1\right)J_0\left(p_2\right)
J_s\left(p_3,z\right)\right\rangle
\right\rangle=\left(1+\left(-1\right)^s\right)i^{s}\pi^{d/2}
\delta\left(\Sigma_k\,p_i\right)
\\&\qquad
\int {d\tau\over\tau^{d/2+1}}\,\tau^3\int \mathrm{d}^3\alpha\,
e^{-\tau\sum\alpha_i\alpha_j\,p_k^2}
f^{(s)}\left(z\cdot \left(\alpha_2p_3-\alpha_3p_2\right)
,z\cdot \left(\alpha_1 p_3-\alpha_3 p_1\right)\right)\,.
\end{split}
\end{equation}
Restoring the normalization, we have
\begin{equation}
\begin{split}
&\left\langle J_0\left(p_1\right)J_0\left(p_2\right)
J_s\left(p_3,z\right)
\right\rangle=
{\left(1+\left(-1\right)^s\right)i^{s}
\over \left[2^{d/2-2}\Gamma\left(\Delta\over 2\right)\right]^3}
\delta\left(\Sigma_k\,p_i\right)
\int {d\tau\over\tau^{d/2+1}}\,\tau^3
\times\\&\qquad\times
\int \mathrm{d}^3\alpha\,
e^{-\tau\sum\alpha_i\alpha_j\,p_k^2}
f^{(s)}\left(z\cdot \left(\alpha_2p_3-\alpha_3p_2\right)
,z\cdot \left(\alpha_1 p_3-\alpha_3 p_1\right)\right)\,.
\end{split}
\end{equation}
The reader may set $s=0$ and compare the resulting expression
with Equation (3.2) of \cite{Gopakumar:2003ns}. We explicitly see
that the parameter $\tau$ constructed from the Schwinger parameters
as above plays the role of the worldline modulus in a first quantized
formulation of the CFT. This connection
was further developed on in \cite{Gopakumar:2004qb}.
\subsection{Fourier Transforming to Position Space}
\label{app:momentum to position}
Having obtained the correlator in momentum space we now Fourier
transform it to position space via
\begin{equation}
\left\langle\left\langle J_0\left(x_1\right)J_0\left(x_2\right)
J_s\left(x_3,z\right)
\right\rangle\right\rangle
=\int {d^dp_1\,d^dp_2\,d^dp_3\over\left(2\pi\right)^{3d/2}} 
e^{i\,p_i\cdot x_i} \left\langle
\left\langle J_0\left(p_1\right)J_0\left(p_2\right)
J_s\left(p_3,z\right)\right\rangle
\right\rangle\,.
\end{equation}
Using the Fourier representation of the Dirac delta function,
\begin{equation}
\begin{split}
\left\langle\left\langle J_0\left(p_1\right)J_0\left(p_2\right)
J_s\left(p_3,z\right)\right\rangle\right\rangle
=&i^s\pi^{d/2}
\left(1+\left(-1\right)^s\right)
\int {d^dw\over\left(2\pi\right)^d}\,e^{-i\left(p_1+p_2+p_3\right)\cdot w}
\times\\
\times
\int {d\tau\over\tau^{d/2+1}}\,\tau^3\,\mathrm{d}^3\alpha\,
f^{(s)}&\left(z\cdot \left(\alpha_1p_3-\alpha_3p_1\right),
z\cdot \left(\alpha_2p_3-\alpha_3p_2\right)\right)
e^{-\tau\sum \alpha_i\alpha_j\,p_k^2}\,.
\end{split}
\end{equation}
The basic momentum integral is
\begin{equation}
\int {d^dp_1\, d^dp_2\, d^dp_3\over \left(2\pi\right)^{3d/2}}
e^{i\,p_i\cdot\left(x_i-w\right)}e^{-\tau\sum \alpha_i\alpha_j\,p_k^2}
=
{1\over \left(2 \tau\right)^{3d/2}\left(\alpha_1\alpha_2\alpha_3\right)^d}
e^{-\sum{\left(x_i-w\right)^2\over 4\tau\alpha_j\alpha_k}}\,.
\end{equation}
As a result, since $p=-i\partial$,
\begin{equation}
\begin{split}
&\left\langle\left\langle J_0\left(x_1\right)J_0\left(x_2\right)
J_s\left(x_3,z\right)\right\rangle\right\rangle=
\left(1+\left(-1\right)^s\right){\pi^{d/2}\over 2^{3d/2}}
\int {d^dw\over \left(2\pi\right)^d}
\int {d\tau\over\tau^{d/2+1}}\,\tau^3\,\mathrm{d}^3\alpha\,
\times\\
&\qquad\times f^{(s)}\left(z\cdot \left(\alpha_1\partial_3-\alpha_3\partial_1\right),
z\cdot \left(\alpha_2\partial_3-\alpha_3\partial_2\right)\right)
{e^{-\sum{\left(x_i-w\right)^2\over 4\tau\alpha_j\alpha_k}}
\over \tau^{3d/2}\left(\alpha_1\alpha_2\alpha_3\right)^d}
\,.
\end{split}
\end{equation}
We finally obtain, also using the fact that $z^2=0$,
\begin{equation}
\begin{split}
f^{(s)}&\left(z\cdot \left(\alpha_1\partial_3-\alpha_3\partial_1\right),
z\cdot \left(\alpha_2\partial_3-\alpha_3\partial_2\right)\right)
\tfrac{e^{-\sum{\left(x_i-w\right)^2\over 4\tau\alpha_j\alpha_k}}}
{\tau^{3d/2}\left(\alpha_1\alpha_2\alpha_3\right)^d}\\
&\qquad=
f^{(s)}\left( \tfrac{z\cdot x_{13}}{2\tau\alpha_2},
\tfrac{z\cdot x_{23}}{2\tau\alpha_1}\right)
\tfrac{e^{-\sum{\left(x_i-w\right)^2\over 4\tau\alpha_j\alpha_k}}}
{\tau^{3d/2}\left(\alpha_1\alpha_2\alpha_3\right)^d}\,.
\end{split}
\end{equation}
As a result,
\begin{equation}
\begin{split}
\left\langle\left\langle J_0\left(x_1\right)J_0\left(x_2\right)
J_s\left(x_3,z\right)\right\rangle\right\rangle=
\left(1+\left(-1\right)^s\right)&{\pi^{d/2}\over 2^{3d/2}}
\int {d^dw\over \left(2\pi\right)^d} 
\int {d\tau\over\tau^{d/2+1}}\,\tau^3\,\mathrm{d}^3\alpha
\times
\\&\times
f^{(s)}\left(\tfrac{z\cdot x_{13}}{2\tau\alpha_2},
\tfrac{z\cdot x_{23}}{2\tau\alpha_1}\right)
\tfrac{e^{-\sum{\left(x_i-w\right)^2\over 4\tau\alpha_j\alpha_k}}}
{\tau^{3d/2}\left(\alpha_1\alpha_2\alpha_3\right)^d}\,.
\end{split}
\end{equation}
Comparing with \eqref{doublebracket00s} we see that this implies
that the correlator is given by
\begin{equation}\label{cft3pt00s app}
\begin{split}
\left\langle J_0\left(x_1\right)J_0\left(x_2\right)
J_s\left(x_3,z\right)\right\rangle=
\left(1+\left(-1\right)^s\right)
&{\pi^{d/2}\over 2^{3\Delta}\Gamma\left(\Delta\over 2\right)^3}
\int {d^dw\over \left(2\pi\right)^d} 
\int {d\tau\over\tau^{d/2+1}}\,\tau^3\,\mathrm{d}^3\alpha
\times
\\&\times
f^{(s)}\left(\tfrac{z\cdot x_{13}}{2\tau\alpha_2},
\tfrac{z\cdot x_{23}}{2\tau\alpha_1}\right)
\tfrac{e^{-\sum{\left(x_i-w\right)^2\over 4\tau\alpha_j\alpha_k}}}
{\tau^{3d/2}\left(\alpha_1\alpha_2\alpha_3\right)^d}\,.
\end{split}
\end{equation}

\subsection{Writing in Embedding Space}
We now multiply the right hand side with $``\,1= {1\over\Gamma\left(s+d-3\right)}
\int_0^\infty {d\rho\over\rho}\,\rho^{s+d-3}\,e^{-\rho}\," $, and
carry out the change of variables proposed in \cite{Gopakumar:2003ns}, namely,
\begin{equation}
t=4\,\tau\,\rho\,\alpha_1\alpha_2\alpha_3\,.
\end{equation}
It is quite straightforward to find the resulting expression
\begin{equation}
\begin{split}
&\left\langle J_0\left(x_1\right)J_0\left(x_2\right)
J_s\left(x_3,z\right)\right\rangle=
\int d^dw
\int {dt\over t^{d/2+1}}\,{t^3\over t^{3d/2}}
\int d\rho\,\rho^2\,\mathrm{d}^3\alpha\,
e^{-\rho}\times \\&\qquad\times
f^{(s)}\left(\tfrac{2\rho\alpha_1\,\rho\alpha_3}{t}\,z\cdot x_{13},
\tfrac{2\rho\alpha_2\,\rho\alpha_3}{t}\,z\cdot x_{23}\right)
e^{-\sum{\rho\alpha_i\over t}\,\left(x_i-w\right)^2}
\left(\rho\alpha_1\,\rho\alpha_2\,\rho\alpha_3\right)^{d-3}\,.
\end{split}
\end{equation}
Next, change variables to $\rho_i=\rho\alpha_i$ to obtain
\begin{equation}
\begin{split}
\left\langle J_0\left(x_1\right)J_0\left(x_2\right)
J_s\left(x_3,z\right)\right\rangle&=
\int d^dw
\int {dt\over t^{d/2+1}}\,{t^3\over t^{3d/2}}\int d^3\rho\,
e^{-\sum_i\left(\rho_i+{\rho_i\over t}\,\left(x_i-w\right)^2\right)}\times \\&\qquad\times
f^{(s)}\left(\tfrac{2\rho_1\,\rho_3}{t}\,z\cdot x_{13},
\tfrac{2\rho_2\,\rho_3}{t}\,z\cdot x_{23}\right)
\left(\rho_1\,\rho_2\,\rho_3\right)^{d-3}\,.
\end{split}
\end{equation}
We next rescale $\rho_i$ to $\sqrt{t}\rho_i$ to obtain
\footnote{
The reader would readily recognize the $s=0$ case,
namely,
\begin{equation}\label{scalar gluing}
\begin{split}
\left\langle J_0\left(x_1\right)J_0\left(x_2\right)
J_0\left(x_3\right)\right\rangle &=
\int d^dw
{dt\over t^{d/2+1}}\int d^3\rho\,
\prod_{i=1}^3 \rho_i^{d-3}
e^{-\rho_i\left(\sqrt{t}+{\left(x_i-w\right)^2\over \sqrt{t}}\right)}
\\&=\int d^dw
{dt\over t^{d/2+1}}\int d^3\rho\,
\prod_{i=1}^3 K_{d-2}\left(x_i;t,z\right)\,,
\end{split}
\end{equation}
which is the integrated product of three bulk to boundary
propagators of $\Delta=d-2$ scalars in AdS$_{d+1}$. This is of course the
essential computation of \cite{Gopakumar:2003ns}.
}
\begin{equation}\label{cft 3pt intrinsic}
\begin{split}
\left\langle J_0\left(x_1\right)J_0\left(x_2\right)
J_s\left(x_3,z\right)\right\rangle=
2^s\int d^dw
{dt\over t^{d/2+1}}\int d^3\rho\,
f^{(s)}&\left(\rho_1\,\rho_3\,z\cdot x_{13},
\rho_2\,\rho_3\,z\cdot x_{23}\right)
\times\\&\times
\prod_{i=1}^3 \rho_i^{d-3}
e^{-\rho_i\left(\sqrt{t}+{\left(x_i-w\right)^2\over \sqrt{t}}\right)}\,.
\end{split}
\end{equation}
%where we dropped a $2^s$. 
We would now like to argue that this expression represents the spin 00$s$
amplitude in AdS$_{d+1}$. As remarked previously, while this match was 
relatively clean for the $s=0$
case, for spinning fields there are potential subtleties. 
It is far from obvious how the expression \eqref{cft 3pt intrinsic}
may be organized into an expectedly complicated bulk expression involving
AdS covariant derivatives and a gauge dependent bulk to boundary propagator.

As remarked previously, it is convenient to re-express \eqref{cft 3pt intrinsic}
in $d+2$ dimensional embedding space $\mathbb{M}^{1,d+1}$ 
of signature $\left(-+\ldots+\right)$ in order to carry out the match with AdS.
Define the points 
\begin{equation}\label{boundary embedding}
P_i=\left(\tfrac{1+x_i^2}{2},\vec{x}_i,\tfrac{1-x_i^2}{2}\right)\,,
\quad W = \left(\tfrac{1+w^2}{2},\vec{w},\tfrac{1-w^2}{2}\right)\,.
\end{equation}
As a result, 
\begin{equation}
\left(x_i-w\right)^2 = -2\,P_i\cdot W\,.
\end{equation}
These are the embedding space representatives of the boundary points
$x_i$ and $w$ respectively.
Also, we define  
\begin{equation}
X=\left(\tfrac{t+w^2+1}{2 t^{1/2} },\tfrac{\vec{w}}{t^{1/2}}, \tfrac{1-t-w^2}{2 t^{1/2} }\right)
=t^{-1/2}\,W+{t^{1/2}\over 2}\left(1,\vec{0},-1\right)\,,
\end{equation}
and hence
\begin{equation}
-2 P_i\cdot X = -2t^{-1/2} P_i\cdot W+t^{1/2}
=\sqrt{t}+{\left(x_i-w\right)^2\over\sqrt{t}}\,.
\end{equation}
The reader would recognize $X$ as a point in AdS$_{d+1}$ embedded 
as the locus $X^2=-1$ in $\mathbb{M}^{1,d+1}$.
The embedding space representative of the intrinsic polarization vector $z$
is denoted by $Z$ and is defined to obey the constraints
via $Z^2=0$ and $Z\cdot P_3=0$. A canonical choice is 
\begin{equation}
Z_c = {{\partial\,P_3\over\partial x_3}}\,\cdot z
=\left(\vec{x}_3\cdot \vec{z}, \vec{z}, -\vec{x}_3\cdot \vec{z}\right)\,,
\end{equation}
As a result, for $i=1,2$,
\begin{equation}
Z_c\cdot P_{i}
=\vec{z}\cdot\vec{x}_{i\,3}\,.
\end{equation}
Hence, for a generic choice of polarization vector 
$Z=Z_{c}+\alpha\, P_3$ we would have
\begin{equation}
z\cdot x_{i3}= Z\cdot P_i-\alpha\, P_i\cdot P_3\,.
\end{equation}
Choosing $\alpha = {\left(Z\cdot X\right)
\over \left(P_3\cdot X\right)}$, we have
\begin{equation}
z\cdot x_{i3}= Z\cdot P_i- 
{\left(Z\cdot X\right)\over \left(P_3\cdot X\right)} P_i\cdot P_3\,.
\end{equation}
With these inputs, and defining $\Delta=d-2$, \eqref{cft 3pt intrinsic} reduces to
\begin{equation}\label{cft3pt embed}
\begin{split}
&\left\langle J_0\left(P_1\right)J_0\left(P_2\right)
J_s\left(P_3,Z\right)\right\rangle=
\sum_{k=0}^s{\left(-1\right)^k 
\over k!\left(s-k\right)!
\left(k+\tfrac{d-4}{2}\right)!
\left(s-k+\tfrac{d-4}{2}\right)!}
\int_{\text{AdS}} d^{d+1}\mathbf{x}
\\&\qquad\times
\int {d\rho_1\over\rho_1}{d\rho_2\over\rho_2}{d\rho_3\over\rho_3}\,
\rho_1^{\Delta+k}\rho_2^{\Delta+s-k}
\rho_3^{\Delta+s}
\left(Z\cdot P_1\,P_3\cdot X 
-Z\cdot X\,P_3\cdot P_1\right)^k
\\&\qquad\qquad\times
\left(Z\cdot P_2\,P_3\cdot X 
-Z\cdot X\,P_3\cdot P_2\right)^{s-k}{1\over \left(P_3\cdot X\right)^s}
\prod_{i=1}^3\,e^{2\,\rho_iP_i\cdot X}\,.
\end{split}
\end{equation}
The reader familiar with embedding space expressions of AdS
propagators \cite{Costa:2014kfa} might already recognize a structure reminiscent
of two scalar bulk to boundary propagators multiplied to a spinning bulk to boundary
propagator, i.e. an AdS amplitude.
\section{The AdS Amplitude in Embedding Space}\label{sec:ads}
Having organized the CFT expression into a form comparable to an AdS amplitude, we now
turn to the AdS computation. The on-shell coupling of a spin-$s$ field to two scalars
is given in intrinsic AdS coordinates $x$ as
\begin{equation}
\hat{\mathcal{V}}_{0,0,s} = g\,\varphi_{\mu_1\ldots\mu_s}\,\varphi^{(a)}
\nabla^{\mu_1}\ldots\nabla^{\mu_s}\varphi^{(b)} = g\,s!\,\varphi^{(a)}\left(\mathbf{x}\right)
\left(\partial_u\cdot\nabla\right)^s\varphi^{(b)}\left(\mathbf{x}\right)
\varphi_{s}\left(\mathbf{x},u\right)\,,
\end{equation}
which in embedding space may be re-expressed as
\begin{equation}\label{onshell embedding}
\hat{\mathcal{V}}_{0,0,s} =g\,s!\,\varphi^{(a)}\left(X\right)
\left(\partial_U\cdot\partial_X\right)^s\varphi^{(b)}\left(X\right)
\varphi_{s}\left(X,U\right)\,,
\end{equation}
The coupling of two scalars to a spin-$s$ field, invariant under the gauge transformation
$\varphi\left(x\right)\rightarrow \varphi\left(x\right)+\nabla \xi\left(x\right)$,
is given by
\begin{equation}\label{intrinsic coupling}
\mathcal{V}_{00s} =\int d^{d+1}\mathbf{x}\,J^{\mu_1\ldots\mu_s}\left(\mathbf{x}\right)
\cdot \varphi_{\mu_1\ldots\mu_s}\left(\mathbf{x}\right)\,.
\end{equation}
where $J^{(s)}$ is a conserved current
constructed from the scalars. 
A specific example of such a current may be constructed in embedding space,
\footnote{Conserved currents in various AdS dimensions were also constructed in
\cite{Prokushkin:1999xq,Prokushkin:1999ke,Fotopoulos:2007yq,Manvelyan:2004mb} and for a conformally coupled scalar
in general dimensions in \cite{Manvelyan:2009tf}. The embedding space formalism allows us to work
with any scalar field in arbitrary dimensions, as pointed out in \cite{Bekaert:2010hk}.
}
leading to the interaction \cite{Bekaert:2010hk}
\begin{equation}\label{v00s}
\mathcal{V}_{00s} = g_s\,s!\,i^s\,\sum_{k=0}^s\,\left(-1\right)^k{s!\over k!\left(s-k\right)!}
\left(\partial_U\cdot\partial_X\right)^k\varphi^{\dagger}\left(X\right)
\left(\partial_U\cdot\partial_X\right)^{s-k}\varphi\left(X\right)\,\Phi\left(X,U\right)\,.
\end{equation}
Since the spin-$s$ field is transverse traceless, we may 
%use transversality, i.e.
%\begin{equation}
%\left(\partial_U\cdot\partial_X\right)\Phi\left(X,U\right)=0\,,
%\end{equation}
%and 
integrate \eqref{v00s} by parts to obtain 
\eqref{onshell embedding} where the respective couplings
are related by $g=g_s\,i^s\,2^s$.

To match with the CFT expression \eqref{cft3pt embed} we shall
consider instead the following current in embedding space
\footnote{As outlined in Appendix \ref{app:embedding},
any field of the form \eqref{embedding current}
\begin{equation}
J\left(X,U\right) = \sum_{k=0}^s\,t_{s,k}\,
\left(U\cdot \partial_X\right)^{k}\varphi^{\dagger}
\left(U\cdot \partial_X\right)^{s-k}\varphi
\end{equation}
obeys the embedding space as well as AdS covariant conservation equation, 
provided the homogeneties of $\varphi\left(X\right)$ and 
$\varphi^{\dagger}\left(X\right)$ are chosen appropriately. 
All such conserved currents of a given spin
in embedding space would differ from each other by longitudinal and
trace terms. Since the spin-$s$
field they are coupling to is transverse traceless, 
this distinction should be irrelevant.
%In case we would also like to ensure that 
%this represents a current in the embedded AdS space
%we also require
%\begin{equation}
%\left(X\cdot\partial_X+U\cdot\partial_U+d\right)
%J=0\,,
%\end{equation}
%given the property that the ambient representative
%of the bulk scalar also satisfies
%\begin{equation}
%\left(X\cdot\partial_X-\left(d-2\right)\right)
%\varphi\left(X\right)=0\,.
%\end{equation}
}
\begin{equation}\label{jtilde}
\tilde{J}_{s} = \sum_{k=0}^s\,
{\left(-1\right)^k\over k!\left(s-k\right)!\left(k+\tfrac{d-4}{2}\right)!\left(s-k+\tfrac{d-4}{2}\right)!}
\left(U\cdot\partial_X\right)^k\varphi^{\dagger}\left(X\right)
\left(U\cdot\partial_X\right)^{s-k}\varphi\left(X\right)\,,
\end{equation}
which we couple to the spin-$s$ field via
\footnote{We may again integrate this vertex by parts, and use the transversality condition
to obtain \eqref{onshell embedding} where the respective couplings
are now related by $g=g_s\, \frac{2^{\Delta +2 s-2} \Gamma 
\left(\frac{1}{2} (2 s+\Delta -1)\right)}{\sqrt{\pi } 
\Gamma (s+1) \Gamma (s+\Delta -1) \Gamma \left(\frac{1}{2} 
(2 s+\Delta )\right)}$\,.
}
\begin{equation}
\mathcal{V}_{00s} = g_s\,\Phi\left(X,\partial_U\right)\,
\sum_{k=0}^s\,
\tfrac{\left(-1\right)^k}{ k!\left(s-k\right)!
\left(k+\tfrac{d-4}{2}\right)!\left(s-k+\tfrac{d-4}{2}\right)!}
\left(U\cdot\partial_X\right)^k\varphi^*\left(X\right)
\left(U\cdot\partial_X\right)^{s-k}\varphi\left(X\right)\,.
\end{equation}
Computing the corresponding amplitude,
\begin{equation}
\begin{split}
\mathcal{A}_{00s} = g_s\,\int d^{d+1}X\,&
\Pi_s\left(P_3,Z;X,\partial_U\right)\,
\sum_{k=0}^s\,
\tfrac{\left(-1\right)^k}{ k!\left(s-k\right)!
\left(k+\tfrac{d-4}{2}\right)!\left(s-k+\tfrac{d-4}{2}\right)!}\times
\\&\times
\left(U\cdot\partial_X\right)^k
\Pi\left(P_1,X\right)
\left(U\cdot\partial_X\right)^{s-k}
\Pi\left(P_2,X\right)\,.
\end{split}
\end{equation}
To proceed further, we need explicit expressions of bulk to boundary
propagators in embedding space. These were obtained in \cite{Costa:2014kfa},
to which we refer the reader for details.
\begin{equation}
\Pi_{\Delta_s,s}\left(P_3,Z;X,U\right)
={C_{\Delta_s,s}}
{\left[2\left(U\cdot P_3\right)
\left(X\cdot Z\right)-2\left(U\cdot Z\right)
\left(X\cdot P_3\right)\right]^s
\over\left(-2 P_3\cdot X\right)^{\Delta_s+s}}\,,
\end{equation}
here $\Delta_s$ is the dimension of the spin-$s$ field, i.e. $\Delta_s=\Delta+s=d-2+s$. Also
\begin{equation}
{C_{\Delta_s,s}}
={\left(s+\Delta_s-1\right)\Gamma\left(\Delta_s\right)
\over 2\pi^{d/2}\left(\Delta_s-1\right)
\Gamma\left(\Delta_s+1-{d\over 2}\right)}\,.
\end{equation}
One may explicitly convert this expression to intrinsic coordinates
and recognize that this is the spin-$s$ bulk to boundary propagator 
in the transverse-traceless gauge \cite{Mikhailov:2002bp}.
We refer the reader to \cite{Sleight:2016hyl} for a review of propagators in embedding space
for different gauge choices. 
Further, using a Schwinger parametrization
for $1\over\left(-2 P_3\cdot X\right)^{\Delta_s}$, 
and identifying $\Delta_s=\Delta+s$, we
may write
\begin{equation}
\Pi_{\Delta_s,s}\left(P_3,Z;X,U\right)
={C_{\Delta+s,s}\over\Gamma\left(\Delta+s\right)}
{\left[\left(2\, P_3\,
\left(X\cdot Z\right)-2\,Z
\left(X\cdot P_3\right)\right)\cdot U\right]^s
\over\left(-2 P_3\cdot X\right)^{s}}\,
\int_0^\infty {d\rho\over\rho}\,\rho^{\Delta+s}\,
e^{-2\,\rho\,P_3\cdot X}\,.
\end{equation}
The amplitude then evaluates to
\begin{equation}
\begin{split}
\mathcal{A}_{00s} = 
&{2^s\,g_s\over\Gamma\left(\Delta\right)^2}
\int d^{d+1}X\,
\Pi_s\left(P_3,Z;X,\partial_U\right)\,
\sum_{k=0}^s\,
\tfrac{\left(-1\right)^k}{ k!\left(s-k\right)!
\left(k+\tfrac{d-4}{2}\right)!\left(s-k+\tfrac{d-4}{2}\right)!}\times
\\&\quad\times
\int{d\rho_1\over\rho_1}{d\rho_2\over\rho_2}
\rho_1^{\Delta+k}\rho_2^{\Delta+s-k}
\left(U\cdot P_1\right)^k
\left(U\cdot P_2\right)^{s-k}
e^{2\rho_1\,P_1\cdot X}e^{2\rho_2\,P_2\cdot X}\,.
\end{split}
\end{equation}
%We checked on Mathematica that 
%\begin{equation}
%\left(V\cdot \partial_U\right)^s \,
%\left(U\cdot P_1\right)^k
%\left(U\cdot P_2\right)^{s-k} = 
%s!\left(V\cdot P_1\right)^k
%\left(V\cdot P_2\right)^{s-k}\,,
%\end{equation}
%provided $V^2=0$.
%We can rewrite the above as
%\begin{equation}
%\Pi_{\Delta_s,s}\left(P_3,Z;X,U\right)
%={C_{\Delta_s,s}}
%{\left[\left(2\, P_3\,
%\left(X\cdot Z\right)-2\,Z
%\left(X\cdot P_3\right)\right)\cdot U\right]^s
%\over\left(-2 P_3\cdot X\right)^{\Delta_s+s}}\,,
%\end{equation}
%from which we see that $V=2\, P_3\,
%\left(X\cdot Z\right)-2\,Z
%\left(X\cdot P_3\right)$ satisfies
%$V^2=0$. Thus we see that the spin-$s$ field
%couples to the traceless part of the current \eqref{jtilde}.
As a result,
\begin{equation}
\begin{split}
\mathcal{A}_{00s} = 
g_s\,&{2^s\,s!\over\Gamma\left(\Delta\right)^2}\,
{C_{\Delta+s,s}\over\Gamma\left(\Delta+s\right)}
\sum_{k=0}^s\,
\tfrac{\left(-1\right)^k}{ k!\left(s-k\right)!
\left(k+\tfrac{d-4}{2}\right)!\left(s-k+\tfrac{d-4}{2}\right)!}
\int d^{d+1}X\,{1\over\left(-2 P_3\cdot X\right)^{s}}
\,
\\&\times
\int{d\rho_1\over\rho_1}{d\rho_2\over\rho_2}
{d\rho_3\over\rho_3}
\rho_1^{\Delta+k}\rho_2^{\Delta+s-k}
\rho_3^{\Delta+s}
\left[\left(2\, P_3\,
\left(X\cdot Z\right)-2\,Z
\left(X\cdot P_3\right)\right)\cdot P_1\right]^k
\\&
\left[\left(2\, P_3\,
\left(X\cdot Z\right)-2\,Z
\left(X\cdot P_3\right)\right)\cdot P_2\right]^{s-k}
e^{2\rho_1\,P_1\cdot X}e^{2\rho_2\,P_2\cdot X}
e^{2\rho_3\,P_3\cdot X}\,.
\end{split}
\end{equation}
This further evaluates to
\begin{equation}
\begin{split}
\mathcal{A}_{00s} = &g_s\,{8^s\,s!\over\Gamma\left(\Delta\right)^2}\,
{C_{\Delta+s,s}\over\Gamma\left(\Delta+s\right)}
\sum_{k=0}^s\,
\tfrac{\left(-1\right)^k}{ k!\left(s-k\right)!
\left(k+\tfrac{d-4}{2}\right)!\left(s-k+\tfrac{d-4}{2}\right)!}
\int d^{d+1}X\,{1\over\left(-2 P_3\cdot X\right)^{s}}
\\&\times
\int{d\rho_1\over\rho_1}{d\rho_2\over\rho_2}
{d\rho_3\over\rho_3}
\rho_1^{\Delta+k}\rho_2^{\Delta+s-k}
\rho_3^{\Delta+s}
\left[P_3\cdot P_1\,
X\cdot Z-Z\cdot P_1
X\cdot P_3\right]^k
\times\\&\qquad\qquad\times
\left[P_3\cdot P_2\,
X\cdot Z-Z\cdot P_2
X\cdot P_3\right]^{s-k}
\prod_{i=1}^3
e^{2\rho_i\,P_i\cdot X}\,.
\end{split}
\end{equation}
which is the same as the CFT answer \eqref{cft3pt embed}, up to an overall proportionality.
We therefore see that the CFT expression naturally organizes itself into an AdS amplitude
determined by an interaction vertex of the form \eqref{intrinsic coupling}, i.e.
\begin{equation}
S_{\text{int}} = \int d^{d+1}\mathbf{x}\,J^{\mu_1\ldots\mu_s}\left(\mathbf{x}\right)\,
\varphi_{\mu_1\ldots\mu_s}\left(\mathbf{x}\right)\,,
\end{equation}
which is the gauge invariant three-point interaction of a spin-$s$ field with two scalars.
We now conclude with some final comments.
\section{Conclusions}
In this paper we showed how the approach of \cite{Gopakumar:2003ns} may be used to organize
correlators of spinning currents into dual AdS amplitudes. The computation presented here may be 
taken as a hint that a skeletal version of open-closed string duality continues to operate even
for the higher-spin/CFT dualities considered here, which are non-supersymmetric and have no
obvious string embedding. In fact, applying the prescription of \cite{Gopakumar:2003ns} leads
us quite naturally to the putative AdS amplitude one would evaluate to match 
with the CFT three-point function
as per \cite{Gubser:1998bc,Witten:1998qj}.

At this juncture we would like to point to the intriguing role played by 
the embedding space formalism in our analysis. In the simplest instance of three scalar operators, we have
seen
\begin{equation}\label{scalar glued}
\begin{split}
&\left\langle J_0\left(P_1\right)J_0\left(P_2\right)J_0\left(P_3\right)\right\rangle
=\int d^dw\,{dt\over t^{d/2+1}}\int_0^\infty \prod_{i=1}^3 {d\rho_i\over\rho_i}\,
\left(\rho_1\rho_2\rho_3\right)^{d-2}\,
e^{-\sum\,\rho_i\left(2\,\tfrac{P_i\cdot W}{t^{1/2}}-t^{1/2}\right)}\\
&\quad=\int d^dw\,{dt\over t^{d/2+1}} \prod_{i=1}^3 \int_0^\infty{d\rho_i\over\rho_i}\,
\rho_i^{d-2}\,e^{-2\rho_i\,P_i\cdot X}
=\int d^{d+1}X\,\prod_{i=1}^3\,K_{d-2}\left(P_i,X\right)\,.
\end{split}
\end{equation}
The above equations make completely manifest a central feature of the AdS/CFT duality.
Namely, the fact that both the CFT and the AdS expressions
are the in fact the same. 
The identification of the quantities $t$ and $ \vec{w}$ as Schwinger parameter
and auxiliary point or the coordinates of an 
interior AdS point is a matter of book-keeping and interpretation. If one wishes
to interpret \eqref{scalar glued} as a CFT correlator then the first choice is 
natural, while interpreting it as an AdS amplitude requires the second choice.
The analysis for the spinning correlators illustrates the same phenomenon, with the additional
feature that the choice of boundary polarization vector representative $Z$ 
in embedding space seems to translate to
gauge choice for the spin-$s$ field in the AdS bulk. 
We would like to better understand this point. 

Further, it is interesting that the CFT amplitude when organized into AdS, doesn't just produce
the on-shell interaction vertex, but seems to yield the interaction of a higher-spin field with a 
conserved current,
as required by bulk gauge invariance.
In a sense therefore, the CFT amplitude does contain information about the full gauge invariant coupling 
of spinning fields to scalars. This is another remarkable, and somewhat gratifying, feature of the
above analysis. Hopefully, these results are indicative of a promising starting point to unravel
the mechanics of AdS/CFT further. We also note here the results of \cite{Didenko:2012vh}
where again embedding space played a crucial role in uncovering a correspondence between the 
CFT two-point function and the AdS bulk to boundary propagator, albeit through somewhat different
means than those adopted here. It would be interesting to explore connections between the two approaches.

\textit{Note:} While this paper was being readied for submission we learned of
\cite{Bzowski:2020kfw} which makes use of the analogy between Feynman graphs 
and electrical circuits to study conformal field theories in momentum space. 
It would also be of interest to develop our methods further
with a view to explicating the AdS/CFT duality directly in momentum space.
\section*{Acknowledgements}
We would like to thank Euihun Joung, 
Rajesh Gopakumar, Nick Halmagyi and Joao Penedones for
helpful discussions. We would especially like to thank Miguel Costa for several 
helpful 
discussions and comments. The work presented here is supported by the Simons 
Foundation grant 488637 (Simons Collaboration
on the Non-perturbative bootstrap) and the project CERN/FIS-PAR/0019/2017. Centro
de Fisica do Porto is partially funded by the Foundation for Science and Technology of
Portugal (FCT) under the grant UID-04650-FCUP. 
\appendix
\section*{Appendix}
\section{The Standard Form of the CFT Three-Point Function}\label{app:standard}
In this section we reduce the expression \eqref{cft3pt00s app}
to the standard form, of the conformal structrure times a prefactor.
In order to do so, we carry out the $w$ integral using
\begin{equation}
\pi^{d/2}\int {d^dw\over\left(2\pi\right)^d}
\,e^{-\sum{\left(x_i-w\right)^2\over 4\tau\alpha_j\alpha_k}}
= \left(\tau\alpha_1\alpha_2\alpha_3\right)^{d/2}
e^{-\sum{\left(x_j-x_k\right)^2\over 4\tau\alpha_i}}\,,
\end{equation}
and as a result
\begin{equation}
\left\langle J_0\left(x_1\right)J_0\left(x_2\right)
J_s\left(x_3,z\right)\right\rangle={1\over 2^{3\Delta}}
\frac{\left(1+\left(-1\right)^s\right)}{\Gamma\left(\Delta\over 2\right)^3}
\int {d\tau}\,\tau^2\,\mathrm{d}^3\alpha\,
f^{(s)}\left(\tfrac{z\cdot x_{13}}{2\tau\alpha_2},
\tfrac{z\cdot x_{23}}{2\tau\alpha_1}\right)
\prod_{i=1}^3
\tfrac{e^{-{x_{jk}^2\over 4\tau\alpha_i}}}
{\left(\tau\alpha_i\right)^{d/2}}\,.
\end{equation}
Now defining $\tau_i=\tau\alpha_i$ we have
\begin{equation}
\left\langle J_0\left(x_1\right)J_0\left(x_2\right)
J_s\left(x_3,z\right)\right\rangle=
{1\over 2^{s+3\Delta}}
\frac{\left(1+\left(-1\right)^s\right)}
{\Gamma\left(\Delta\over 2\right)^3}
\,\int d^3\tau\,
f^{(s)}\left(\tfrac{z\cdot x_{13}}{\tau_2},
\tfrac{z\cdot x_{23}}{\tau_1}\right)
\tfrac{e^{-\sum{x_{jk}^2\over 4\tau_i}}}
{\left(\tau_1\,\tau_2\,\tau_3\right)^{d/2}}\,.
\end{equation}
The $\tau$ integrals can be done to obtain the expected conformal structure.
To see this, expand
\begin{equation}
\begin{split}
\left\langle J_0\left(x_1\right)J_0\left(x_2\right)
J_s\left(x_3,z\right)\right\rangle=
\frac{1}{2^{s+3\Delta}}&
\frac{\left(1+\left(-1\right)^s\right)}
{ n_s^{1/2}\Gamma\left(\Delta\over 2\right)^3}\,
\int d^3\tau\,\sum_{k=0}^s 
\tfrac{\left(-1\right)^k}{k!\left(s-k\right)!
\left(k+{d-4\over 2}\right)!\left(s-k+{d-4\over 2}\right)!}
\times\\&\times
\left(\tfrac{z\cdot x_{13}}{\tau_2}\right)^k
\left(\tfrac{z\cdot x_{23}}{\tau_1}\right)^{s-k}
\tfrac{e^{-\sum{x_{jk}^2\over 4\tau_i}}}
{\left(\tau_1\,\tau_2\,\tau_3\right)^{d/2}}\,.
\end{split}
\end{equation}
We see that the $4\tau_i$ are the inverses of the usual
Schwinger parameters in terms of which which one would typically write the
the three-point amplitude in position space. This is somewhat unsurprising
since the $\tau_i$ were the usual Schwinger parameters in momentum space.
Finally, we carry out the $\tau_i$ integrals using the
identity
\begin{equation}
\int d\tau\,{1\over \tau^n} e^{-x^2/4\tau}
={4^{n-1}\over \left(x^2\right)^{n-1}}\Gamma\left(n-1\right)\,,
\end{equation}
and subsequently summing over $k$,
we obtain
\begin{equation}
\left\langle J_0\left(x_1\right)J_0\left(x_2\right)
J_s\left(x_3,z\right)\right\rangle=
\tfrac{\left(1+\left(-1\right)^s\right)}{s!\,
\Gamma\left(\Delta\over 2\right)^2}
{2^{s}n_s^{-1/2}
\over \left(x_{12}^2\right)^{\Delta\over 2}
\left(x_{23}^2\right)^{\Delta\over 2}
\left(x_{13}^2\right)^{\Delta\over 2}}
\left[
\frac{z\cdot x_{13}}{x_{13}^2}-
\frac{z\cdot x_{23}}{x_{23}^2}\right]^{s}\,,
\end{equation}
which contains the expected conformal structure.
The prefactor may be evaluated using \eqref{ns},
which states
\begin{equation}
n_s
=
\frac{(-1)^s 2^{\Delta +3 s-2} }
{\sqrt{\pi }\,s!\, \Gamma \left(\frac{\Delta }{2}\right)^{2} 
}
\left[ 
\frac{\Gamma \left(s+\frac{\Delta-1}{2}\right)}{
\Gamma \left(s+\frac{\Delta}{2}\right) \Gamma (s+\Delta -1)}
\right]\,,
\end{equation}
and hence the prefactor is
\begin{equation}
-i^s\left(1+\left(-1\right)^s\right)
2^{\frac{1}{2} (2-s+\Delta)}\pi^{1/4}
\left[ 
\frac{\Gamma \left(s+\frac{\Delta}{2}\right) \Gamma (s+\Delta -1)}
{s!\,\Gamma \left(s+\frac{\Delta-1}{2}\right)
\Gamma \left(\tfrac{\Delta }{2}\right)^{2} }
\right]^{1/2}\,.
\end{equation}
\section{Normalized Traceless Currents}\label{app:current normalization}
In this section we provide some details of how the spin-$s$ currents \eqref{js}
are normalized via \eqref{ns}. We consider the two-point function, where conformal invariance
demands that the result is of the form
\begin{equation}
\left\langle J_s\left(x_1,z_1\right)J_s\left(x_2,z_2\right)
\right\rangle
={g\over \left(x_{12}^2\right)^{\Delta+s}}
\left[z_1\cdot z_2-2{z_1\cdot x_{12}\,z_2\cdot x_{12}\over x_{12}^2}\right]^s\,,
\end{equation}
where we have to normalize the currents so that $g=1$. If we adopt the series expansion
form \eqref{fs} for \eqref{js}, then the simplest way to determine this normalization is to compute the 
coefficient of the term ${\left(z_1\cdot z_2\right)^s\over \left(x_{12}^2\right)^{\Delta+2s}}$.
Start with the two-point function in the Schwinger parametrization
\begin{equation}
\begin{split}
&\left\langle J_s\left(x_1,z_1\right)J_s\left(x_2,z_2\right)\right\rangle
=\sum_{k_1,\,k_2=0}^s\,c^s_{k_1}\,c^s_{k_2}
{1\over\Gamma\left(\tfrac{\Delta}{2}\right)^2}\int_0^\infty 
{du_1\over u_1}{du_2\over u_2}\left(u_1\,u_2\right)^{\Delta/2}\times\\
&\times\left(z_1\cdot\partial_{y_1}\right)^{k_1}
\left(z_1\cdot\partial_{\bar{y}_1}\right)^{s-k_1}
\left(z_2\cdot\partial_{y_2}\right)^{k_2}
\left(z_2\cdot\partial_{\bar{y}_2}\right)^{k_2}
e^{-u_1\left(y_1-\bar{y_2}\right)^2-u_2\left(\bar{y}_1-{y_2}\right)^2}
\vert_{\left(y_i,\bar{y}_i\right)\rightarrow x_i}\,.
\end{split}
\end{equation}
Since $z_i^2=0$, one readily sees that the only terms that contribute to 
$\left(z_1\cdot z_2\right)^s$ come with $k_1=k$, $k_2=s-k$. The relevant term 
evaluates to
\begin{equation}
\begin{split}
\left\langle J_s\left(x_1,z_1\right)J_s\left(x_2,z_2\right)\right\rangle
&={2^s\left(-1\right)^s\over\Gamma\left(\tfrac{\Delta}{2}\right)^2}
{\left(z_1\cdot z_2\right)^{s}
\over \left(x_{12}^2\right)^{\Delta+2s}}
\sum_{k=0}^s\,\tfrac{1}{
k!\left(s-k\right)!
\left(k+\tfrac{\Delta}{2}-1\right)!
\left(s-k+\tfrac{\Delta}{2}-1\right)!}
+\ldots\,,
\\&=
n_s\,{\left(z_1\cdot z_2\right)^{s}
\over \left(x_{12}^2\right)^{\Delta+2s}}+\ldots
\end{split}
\end{equation}
with $n_s$ as defined in \eqref{ns}\,.
As an example, the explicit form of the normalized traceless spin two
current in $d$ dimensions is 
\begin{equation}
J_2\left(x,z\right)=
\varphi^*\left(x,z\right)
\frac{\left\lbrace\left(z\cdot\overleftarrow{\partial}\right)^2
-\frac{2d}{d-2} z\cdot\overleftarrow{\partial} 
z\cdot\overrightarrow{\partial}+ 
\left(z\cdot\overrightarrow{\partial}\right)^2
\right\rbrace}{2 \sqrt{2} \sqrt{(d-1) d}}
\varphi\left(x,z\right)\,,
\end{equation}
which may be reproduced by expanding \eqref{js}.
\section{Embedding Space for AdS and CFT}
\label{app:embedding}
In this section we provide an overview of the embedding space formalism, which 
dates back to Dirac, and was further developed by Fronsdal \cite{Fronsdal:1978vb}. The essential
idea is to consider a $d+2$ dimensional Minkowski space $\mathbb{M}^{1,d+1}$ of metric
signature $\left(-,+,\ldots,+\right)$. The $so(d+1,1)$ symmetry acts linearly on this
space, which is the great advantage of this formalism.
Vectors in $\mathbb{M}^{1,d+1}$ carry capital
latin indices, and the  $so(d+1,1)$ invariant
inner product of two such vectors $X^A$ and $Y^B$ is denoted by
\begin{equation}
X\cdot Y = \eta_{AB}\,X^A\,Y^B\,,\qquad X^2=X\cdot X\,.
\end{equation}
AdS$_{d+1}$, with intrinsic cooordinates $\mathbf{x}^{\mu}$, 
may be embedded into this space as the locus $X^2=-R$. In the main text we shall set
$R=1$. We consider the embedding
\footnote{A concrete choice, for $R=1$, that we use in this paper is
\begin{equation}
X=\left(\frac{t+w^2+1}{2 t^{1/2} },\frac{\vec{w}}{t^{1/2}}, \frac{1-t-w^2}{2 t^{1/2} }\right)\,.
\end{equation}
This is Euclidean AdS$_{d+1}$. Its boundary $\mathbb{R}^d$ is the locus $P^2=0$
parametrized by
\begin{equation}
P_i=\left(\frac{1+x^2}{2},\vec{x},\frac{1-x^2}{2}\right)\,.
\end{equation}
The parametrization is somewhat different if one chooses light-cone
coordinates in embedding space. These choices are of course trivially related to each other.}
\begin{equation}
\mathbf{x}^{\mu}\mapsto X^A\left(\mathbf{x}^{\mu}\right)\,.
\end{equation}
Tensors in AdS$_{d+1}$ are mapped to their embedding space
representatives in $\mathbb{M}^{1,d+1}$ via
\begin{equation}
t_{\mu_1\ldots\mu_s}\left(\mathbf{x}\right)
={\partial X^{A_1}\over\partial \mathbf{x}^{\mu_i}}\ldots
{\partial X^{A_s}\over\partial \mathbf{x}^{\mu_s}}
T_{A_1\ldots A_s}\left(X\right)\,,
\end{equation}
where the tensor $T_{A_1\ldots A_r}$ is taken to satisfy the additional
constraints \cite{Fronsdal:1978vb} of 
\begin{enumerate}
\item \textit{homogeneity} of fixed non-zero degree $k\in\mathbb{C}$,
\begin{equation}
T_{A_1\ldots A_s}\left(\lambda\,X\right)=\lambda^k\,T_{A_1\ldots A_s}\left(X\right)\,,
\, \text{ensured by}\, 
\left(X\cdot\partial_X-k\right)T_{A_1\ldots A_s}\left(X\right)=0\,,
\end{equation}
and,
\item \textit{tangentiality} to the AdS$_{d+1}$ loci of any given $R$, i.e.
\begin{equation}
X^{A_1}\,T_{A_1\ldots A_s}\left(X\right)=0\,.
\end{equation}
\end{enumerate}
The second condition is implemented by the projection operator
\begin{equation}
P_A^B=\delta_A^B-\frac{X_A\,X^B}{X^2}\,.
\end{equation}
It is straightforward to check that for a given vector
$Y^A$, $X\cdot P\left(Y\right)=0$\,. Further, the embedding space representative
of the AdS$_{d+1}$ covariant derivative defined with respect to the Levi-Civita
connection is
\begin{equation}
\nabla\mapsto \mathcal{D} = P\circ\partial\circ P\,.
\end{equation}
An important example of these statements is the scalar field itself \footnote{
We refer the reader to \cite{Bekaert:2010hk} for a complete account
of the results mentioned here, with the reminder that in their conventions, $d$
is the AdS dimension, which for us is $d+1$.}. Using the above
statements, one may show that a scalar field $\varphi\left(\mathbf{x}\right)$ 
in AdS$_{d+1}$
may be represented by $\varphi\left(X\right)$ 
of homogeneity $\mu-{d/2}$ where
\begin{equation}
\partial^2\varphi\left(X\right)=0\,\longleftrightarrow\,
\left[\nabla^2_{AdS_{d+1}}+{1\over R^2}\left({d^2\over 4}-\mu^2\right)\right]
\varphi\left(\mathbf{x}\right)=0\,.
\end{equation}
If we identify this with the AdS$_{d+1}$ Klein Gordon equation
$\left[\nabla^2_{AdS_{d+1}}-m^2\right]\varphi\left(\mathbf{x}\right)=0$
and note the dictionary that 
$\left(m\,R\right)^2=\Delta\left(\Delta-d\right)$
then
\begin{equation}
\mu={d\over 2}-\Delta\,,\Delta-{d\over 2}\qquad
\Rightarrow\qquad k=-\Delta\,,-\left(d-\Delta\right)\,.
\end{equation}
In the following, we shall take the embedding representative 
of $\varphi\left(\mathbf{x}\right)$
to have homogeneity $k=-\Delta$ while the embedding 
representative of the complex conjugate
$\varphi^*\left(\mathbf{x}\right)$ will be a field 
$\varphi^{\dagger}\left(X\right)$ of homogeneity
$k^{\dagger}=\Delta-d$.

Finally, since we shall always work with completely symmetric tensors, we find it
convenient to introduce polarization vectors $u^i\mapsto U^A$ with which we shall
contract all indices. Therefore, a tensor $T_{A_1\ldots A_s}\left(X\right)$
will be written as 
\begin{equation}
T_{s}\left(X,U\right)=T_{A_1\ldots A_s}\left(X\right) U^{A_1}\ldots U^{A_s}\,.
\end{equation}
The subscript will be dropped when there is no danger of ambiguity. 
We now turn to fields in AdS$_{d+1}$ which obey the conservation law
\begin{equation}
\partial_{u}\cdot\nabla\,J_s\left(\mathbf{x},u\right)\approx 0\,.
\end{equation}
To this end, we consider fields in embedding space of the form
\begin{equation}\label{embedding current}
J_{s}\left(X,U\right) = \sum_{m=0}^s t_{s,m}\,
\left[\left(U\cdot\partial_X\right)^m
\varphi^{\dagger}\left(X\right)\right]
\left[
\left(U\cdot\partial_X\right)^{s-m}\varphi\left(X\right)\right]\,,
\end{equation}
where $t_{s,m}$ are arbitrary coefficients.
This clearly obeys the conservation equation
\begin{equation}
\partial_U\cdot\partial_X\,J_{s}\left(X,U\right)\approx 0\,,
\end{equation}
provided $\partial_X^2\varphi=0=\partial_X^2\varphi^{\dagger}$.
This also obeys the covariant conservation law
\begin{equation}
\mathcal{D}_{A_1}\,J^{A_1\ldots A_s}\approx 0\,,
\end{equation}
since \cite{Bekaert:2010hk}, as one may readily check,
\footnote{It is useful to note the identity
\begin{equation}
X\cdot\partial_X\,\left(U\cdot\partial_X\right)^n
=\left(U\cdot\partial_X\right)^n\,\left(X\cdot \partial_X-n\right)\,.
\end{equation}
As a result,
\begin{equation}
\begin{split}
X\cdot\partial_X\,&\left(U\cdot\partial_X\right)^m
\varphi^{\dagger}\left(X\right)
\left(U\cdot\partial_X\right)^{s-m}\varphi\left(X\right)
\\&=
\left(k+k^{\dagger}-s\right)\left(U\cdot\partial_X\right)^m
\varphi^{\dagger}\left(X\right)
\left(U\cdot\partial_X\right)^{s-m}\varphi\left(X\right)
\\&
=-\left(d+s\right)
\left(U\cdot\partial_X\right)^m
\varphi^{\dagger}\left(X\right)
\left(U\cdot\partial_X\right)^{s-m}\varphi\left(X\right)\,,
\end{split}
\end{equation}
and \eqref{current homogeneity} readily follows.}
\begin{equation}\label{current homogeneity}
\left(X\cdot\partial_X+U\cdot\partial_U+d\right)\,J_{s}\left(X,U\right)=0\,.
\end{equation}
As a result, \eqref{embedding current} also 
defines a conserved current in 
AdS$_{d+1}$.

The $\mathbb{R}^{d}$ boundary, with intrinsic coordinates $\vec{x}$, is parametrized in embedding space by points $P$ obtained
from the AdS$_{d+1}$ locus $X$ in the scaling limit
\begin{equation}
P^A=\epsilon \,X^A\, \qquad \epsilon\rightarrow 0\,,
\end{equation}
while keeping $P^+=P^0+P^{d+1}$ fixed. 
Since $X^2$ is fixed, this limit sets $P^2=0$. The embedding space representative 
of a traceless boundary tensor $f_{i_1\ldots i_s}$, denoted by $F_{A_1\ldots A_s}$ 
obeys 
\begin{equation}
\eta^{A_1A_2}F_{A_1\ldots A_s}=0\,,
\end{equation}
along with
the usual criteria of homogeneity and tangentiality as in the AdS$_{d+1}$
case. However, since $P^2=0$, the tangentiality criterion
\begin{equation}
P^{A_1}\,F_{A_1\ldots A_s}=0\,,
\end{equation}
leaves a residual ambiguity
\begin{equation}
F_{A_1\ldots A_s}\sim F_{A_1\ldots A_s}+P_{(A_1}G_{A_2\ldots A_s)}\,.
\end{equation}
Again, as we are working with completely symmetric traceless tensors it
is convenient to introduce polarization vectors $z$ and $Z$ in intrinsic
and embedding spaces respectively. It is relatively straightforward to show
that the above properties of the embedding space representative are satisfied
if $Z$ obeys the conditions
\begin{equation}
Z^2=0\,,\qquad Z\cdot P=0\,.
\end{equation}
Then a traceless spin-$s$
field $j_s\left(x,z\right)$ maps to the embedding space representative $J_s\left(P,Z\right)$. Also, 
we may shift the polarization vection $Z$ to $\tilde{Z}= Z+\alpha\,P$, which
also obeys the properties
\begin{equation}
\tilde{Z}^2=0\,,\qquad \tilde{Z}\cdot P=0\,,
\end{equation}
Hence
$J_s\left(P,Z+\alpha\,P\right)$\,,
is an equally good representative of 
$j_s\left(x,z\right)$. This fact is useful in the computations presented in the main text.
Further details about this formalism for CFTs may be found in \cite{Costa:2011mg}.
\bibliography{biblio}
\bibliographystyle{utphys} 

\end{document}